\documentclass[twocolumn,superscriptaddress,showpacs,amsmath,amssymb]{revtex4-1}
\usepackage{amsmath}
\usepackage{amssymb}

\usepackage{graphicx} %Include figure files
\usepackage{dcolumn}% Align table columns on decimal point
\usepackage{txfonts}% bold math
\usepackage[pdfpagemode=UseNone,pdfstartview=FitH,colorlinks=true,linkcolor=blue,urlcolor=blue,anchorcolor=blue,citecolor=blue]{hyperref}
\allowdisplaybreaks[4]

\usepackage{appendix}

\usepackage{color}
\usepackage{soul,xcolor}
\usepackage{ulem}
\setstcolor{red}

\def\pra#1{{ Phys.\ Rev. A\/} {\bf#1}}
\def\prb#1{{ Phys.\ Rev. B\/} {\bf#1}}

\def\prl#1{{ Phys.\ Rev.\ Lett.} {\bf#1}}

\def\sci#1{{ Science} {\bf#1}}

\def\rmp#1{{ Rev. \ Mod. \ Phys.} {\bf#1}}
\def\nat#1{{ Nature} {\bf#1}}

\def\nap#1{{ Nat. Phys.} {\bf#1}}

\begin{document}
\title{ Adiabatic topological pumping  in a  semiconductor nanowire  }
%\title{Quantum spin hall effect on a freetanding InSb nanolayer    }
\author{Zhi-Hai Liu }
\email{zhihailiu@pku.edu.cn}
\affiliation{Beijing Key Laboratory of Quantum Devices, Key Laboratory for the Physics and Chemistry of Nanodevices, and Department of Electronics, Peking University, Beijing 100871, China}
%\affiliation{Beijing Academy of Quantum Information Sciences, Beijing 100193, China}

\author{H.~Q. Xu}
\email{hqxu@pku.edu.cn}
\affiliation{Beijing Key Laboratory of Quantum Devices, Key Laboratory for the Physics and Chemistry of Nanodevices, and Department of Electronics, Peking University, Beijing 100871, China}
\affiliation{Beijing Academy of Quantum Information Sciences, Beijing 100193, China}

\begin{abstract}
The  adiabatic topological pumping is proposed by  periodically modulating  a semiconductor nanowire double-quantum-dot chain.    We demonstrate that  the quantized charge transport can be achieved by a nontrivial modulation of the quantum-dot well  and   barrier potentials.  When the quantum-dot well potential is replaced by  a time-dependent  staggered magnetic field,  the topological spin pumping can be realized by periodically modulating the barrier potentials and magnetic field. We also  demonstrate that in the presence of Rashba spin-orbit interaction, the double-quantum-dot chain can be used to implement the topological spin pumping. However, the pumped spin in this case can have a quantization axis other than the applied magnetic field direction. Moreover, we show that all the adiabatic topological pumping are manifested by the presence of  gapless edge states traversing the band gap as a function of time.

\end{abstract}
%, due to the opposite directions of the  parametric trajectories for two different spin states---logical revion and --the topological charge---by mapping

\date{\today}
\maketitle
\section{Introduction}

In recent years adiabatic topological pumping  has attracted considerable attentions as it plays an important role in implementing  quantized  particle transports~\cite{Meidan2010,Citro2016} and in exploring higher-dimensional topological effects~\cite{Prodan2015,Kraus2016,Zilberberg2018,Lohse2018}. The concept of topological  pumping was first introduced  by Thouless who showed that the quantization of charge transport can be realized by a cyclic modulation of a one-dimensional periodic potential adiabatically and the charge pumped per cycle can be  determined by the Chern number defined over the two-dimensional Brillouin zone formed in the momentum and time spaces~\cite{Thouless1983,Niu1984}. It has been confirmed  that some low-dimensional systems, such as quasicrystals and optical supperlattices, can exhibit the higher-dimensional topological effects, when subjected to a nontrivial modulation, and can thus be used to implement the adiabatic topological pumping ~~\cite{Tangpanitanon2017,Ke2020,Lin2020,Kraus2012,Kraus2013,Verbin2015,Schweizer2016,Nakajima206,Lohse2015}. Recently, the Thouless pumping of ultracold bosonic/fermionic atoms in an optical superlattice  has  been experimentally achieved via  periodically modulating the superlattice phase~\cite{Nakajima206,Lohse2015}. However, up to now, realizations of the topological  pumping  are almost limited to the optical systems and, therefore, it is of interest to explore the exotic phenomena of topological pumping in low-dimensional systems made from conventional semiconductors.

Both experimentally and theoretically,
the adiabatic quantum pumping of electron charge and spin in semiconductor nanostructures has been one of the hotspot issues~\cite{Thomas1994,Brouwer1998,OEA2002,Altshuler1999,Switkes1999,Aleiner1998,Hasegawa2019,OEA2002a,Aharony2002,Zhou1999}.  Especially, the parametric pumping facilitates the adiabatic transfer of noninteracting electrons in unbiased quantum dots~\cite{Brouwer1998,Switkes1999,OEA2002}, but the average transferred charge in a cycle is not necessarily quantized under such a parametric pumping  scheme. It has been shown that the pumped number is quantized through the adiabatic topological pumping and the physics behind this
deeply roots into the nontrivial topology of the periodic wave functions~\cite{Xiao2010,Wang2013,Hatsugai2016}. Moreover, the presence of fractional boundary charges in quantum dot arrays  has been  theoretically demonstrated by manipulating the onsite potentials periodically~\cite{Park2016,Thakurathi2018}. However,  a systematic analysis of the topological pumping in a one-dimensional periodic semiconductor nanostructure  is still scarce.

\begin{figure}
\centering
\includegraphics[width=0.48\textwidth]{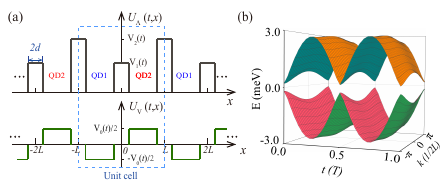}
\caption{(color online)  (a) Schematic diagrams of two periodic potentials $U_{\Lambda}(t,x)$ and $U^{}_{V}(t,x)$ assumed in the  implementation of the topological charge pumping in a semiconductor nanowire DQD chain.  Specifically, $U_{\Lambda}(t,x)$  consists of barrier potentials $V^{}_{1}(t)$ and $V^{}_{2}(t)$ in a unit cell, with $2d$ denoting the width of each barrier potential and $2L$ being the  unit-cell length, and  $U^{}_{V}(t,x)$  comprises two QD well potentials $\pm V^{}_{0}(t)/2$.  (b) 3D  plot of the energy spectrum for the two lowest-energy Bloch bands of the periodic DQD chain with the parameter vector $\widehat{V}=(V_{0},\delta V)$ driven by a nontrivial modulation. Here the parameters used in the calculations are $m^{\ast}_{}=0.023m^{}_{0}$ ($m^{}_{0}$ is the free electron mass), $2L=120$~nm, $2d=20$~nm, $\varepsilon^{}_{0}=6.0$~meV, and the sum of the two barrier potentials $V^{}_{2}(t)+V^{}_{1}(t)=36$~meV.}
\label{Fig1}
\end{figure}

In this paper, we  propose a scheme for implementing the adiabatic topological pumping in a semiconductor nanowire double-quantum-dot (DQD) chain. We first demonstrate  that the topological charge pumping can be achieved by simultaneously modulating the quantum-dot well and barrier potentials of the DQD chain. In the absence of the quantum-dot well potential modulation, we show that the topological spin pumping can be achieved by employing  a time-dependent staggered magnetic field and by a nontrivial modulation of the barrier potentials and magnetic  field in a cycle. When the DQD chain of a finite length is coupled to the external  leads, we  show that the quantized charge and spin transport  can be calculated by exploiting the scattering matrix formalism  under a nontrivial modulation. Especially,  we demonstrate the quantized spin pumping in the presence of the Rashba spin-orbit interaction (SOI), but the pumped spin can have the spin quantization axis different from the applied staggered magnetic field direction. We also demonstrate, based on the changing  of  time-reversal charge polarization  in a half cycle, that the periodic DQD chain in the presence of SOI can serve  as a dynamic version of a topological insulator under a nontrivial modulation of the barrier potentials and magnetic field.

The rest part of the paper is organized as follows. In Sec.~\ref{SECII}, a periodic potential to define the DQD chain in a one-diemnsional semiconductor is introduced and the implementation of topological charge pumping is demonstrated.  In Sec.~\ref{SECIII}, the topological spin pumping in the semiconductor DQD  chain in the presence of  a time-dependent staggered magnetic field is analyzed.  Section~\ref{SECIV} devotes to the formalism for calculating the charge and spin pumping through a finite DQD chain under topological modulations.   In the presence of the Rashba SOI,  the topological spin pumping in the periodic DQD chain is discussed in Sec.~\ref{SECV}. Finally,  we summarize the paper in Sec.~\ref{SECVI}.

%continuous Hamiltonian of  the non-interacting double-quantum-dot chain is introduced, and the

\section{Topological charge pumping}~\label{SECII}

%We consider a  periodic semiconductor nanowire DQD system subjected to a time-dependent  , as shown in,  described by the effective ONE-DIMENSIONAL hAMILTONIAN OF

We first consider a periodic semiconductor nanowire DQD system subjected to time-dependent periodic barrier potential $U^{}_{\Lambda}(t,x)$ and quantum-dot well potential $U^{}_{V}(t,x)$,  as shown in Fig.~\ref{Fig1}(a), described by the effective one-dimensional Hamiltonian,
\begin{align}
H^{}_{0}=\frac{p^{2}}{2m^{\ast}_{}}+U^{}_{\Lambda}(t,x)+U^{}_{V}(t,x)\ ,
\label{H0}
\end{align}
where $p=-i\hbar \frac{\partial}{\partial x}$ represents the momentum operator and $m^{\ast}_{}$ is the effective electron mass. In this system, a unit cell, as indicated by the dashed square in Fig.~\ref{Fig1}(a), comprises two quantum dots (QDs), labeled  as QD1 and QD2.
%Structurally, {\color{red} $U(t,x)$ can be separated into} two terms $U^{}_{}(t,x)=U^{}_{\rm \Lambda}(t,x)+U^{}_{\rm V}(t,x)$,
 Structurally,  $U^{}_{\rm \Lambda}(t,x)$ consists of  barrier potentials $V^{}_{1}(t)$ and $V^{}_{2}(t)$ and    $U^{}_{\rm V}(t,x)$ comprises two QD well  potentials $\pm V^{}_{0}(t)/2$ in a unit cell. Explicitly, for a unit cell of $-L\leq x<L$,   the two potentials can be written as
 \begin{align}
 U^{}_{\rm V}(t,x)= & \frac{1}{2}V^{}_{0}(t) [f^{+}_{1}(x)-f^{+}_{2}(x)],\nonumber  \\
 U^{}_{\rm \Lambda}(t,x)= & V^{}_{1}(t)f^{-}_{1}(x) +V^{}_{2}(t)[1+f^{-}_{2}(x)]\ ,
 \label{CONS}
\end{align}
 where $2L$ denotes the size of unit cell, $2d$ ($2d\ll 2L$)  the  width of each potential barrier,   $f^{\pm}_{1}(x)=\Theta(x+d)\pm \Theta(x-d)$ and $f^{\pm}_{2}(x)=\Theta(x+d-L)\pm \Theta(x+L-d)$,  with  $\Theta(x)$ being the Heaviside function, the spatial distribution functions.

\begin{figure}
\centering
\includegraphics[width=0.485\textwidth]{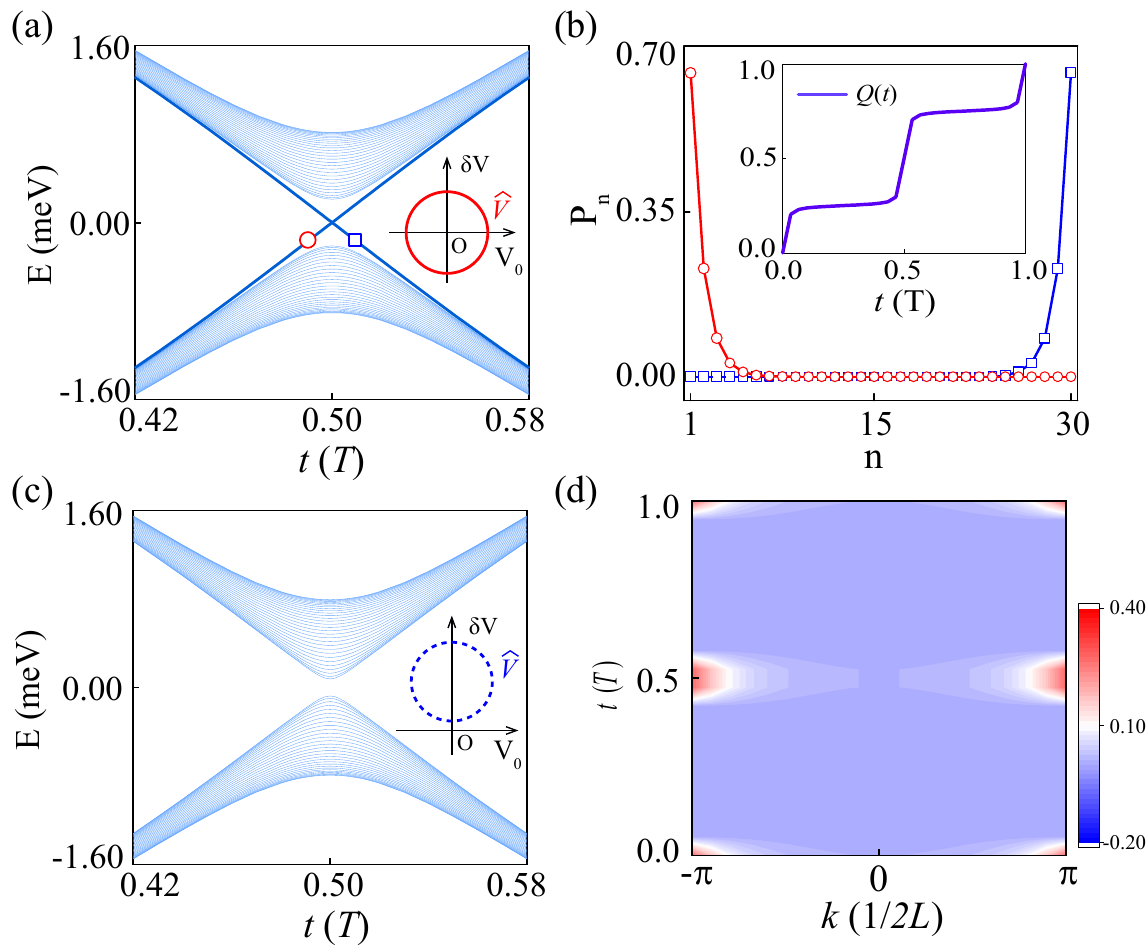}
\caption{(color online) (a) Energy spectrum of a  finite DQD chain comprising 30 unit cells  versus $t$ under a nontrivial modulation of $\widehat{V}=(V^{}_{0},\delta V)$ with  $\varepsilon^{}_{0}=6.0$~meV and a changing contour of $\widehat{V}$ as  displayed in the inset. (b) Probability  density distributions, $P^{}_{n}$, calculated for the two gapless band states of the finite DQD chain, as indicated by the circle and square symbols in (a). Inset shows the calculated time-evolution of charge polarization $\mathcal{Q}(t)$. (c) The same as (a), but under a trivial modulation  with   $\varepsilon^{}_{0}=6.0$~meV and    $\varepsilon^{}_{\iota}=8.0$~meV.  (d) Berry curvature $F(k,t)$ calculated as a function of $k$ and $t$ for the occupied Bloch band of the corresponding infinite DQD chain under the nontrivial modulation as shown in the inset of (a). All other parameters employed are the same as in Fig.~\ref{Fig1}.}
\label{Fig2}
\end{figure}

%and the DQD chain  is equivalent to a tight-binding model with the effective Hamiltonian given by
 In fact,  the topological charge pumping can be achieved by  a nontrivial modulation  of system parameters  $V_{0}(t) $ and  $\delta V(t)\equiv V^{}_{2}(t)-V^{}_{1}(t)$. To see this, we consider the case that there is only one energy level in each QD and rewrite Hamiltonian Eq.~(\ref{H0}) in the second quantization form
\begin{align}
H^{}_{\rm 0,T}=&\sum^{}_{n}\left(t^{}_{\rm in,0}a^{\dagger}_{n}b^{}_{n}+t^{}_{\rm ex,0} a^{\dagger}_{n+1}b^{}_{n}+\rm{h.c.}\right)\nonumber \\
&~~-\frac{\Delta^{}_{0}}{2}\sum_{n}\left( a^{\dagger}_{n}a^{}_{n}-b^{\dagger}_{n}b^{}_{n}\right)\ ,
\label{H0T}
\end{align}
with $n$ representing the unit-cell index, $a^{\dagger}_{n }$ ($a^{}_{n}$) and $b^{\dagger}_{ n}$ ($b^{}_{n}$) denoting  the creation (annihilation) operators of the energy levels in QD1 and QD2 of the $n$th unit cell,  $t^{}_{\rm in/ex,0 }$ being the intra/inter unit-cell hopping amplitudes,  and $\pm\Delta^{}_{0}/2$  the  effective on-site energies (i.e., the level energies) in QD1 and QD2. Note that  the Hamiltonian in Eq.~(\ref{H0T})  is derived from Eq.~(\ref{H0}) by subtracting a constant potential value and
 this Hamiltonian is identical to the Rice-Mele model  Hamiltonian~\cite{Rice1982}.
In analogy with the Thouless pumping in Refs.~\onlinecite{Nakajima206,Lohse2015}, it is found that  the topological charge pumping can be realized when the changing trajectory of the parameter  vector $(\Delta^{}_{0},\delta t^{}_{0})$, with $\delta t^{}_{0} =t^{}_{\rm in,0}  -t^{}_{\rm ex,0} $, obeys a nonzero winding number in a period. Here, because $\Delta^{}_{0}$ and  $\delta t^{}_{0}$ can be regulated by potentials $V_{0}$ and $\delta V$, respectively,  we can deduce that the topological  charge pumping  can be implemented if the changing contour of the modulation vector $\widehat{V}=(V_{0},\delta V)$ possesses a nonzero winding number in a cycle.

For an infinite periodic DQD chain system under a nontrivial modulation with  $V^{}_{0}(t)=\varepsilon^{}_{0}\sin(2\pi t/T )$ and $\delta V(t)=\varepsilon_{0}\cos(2\pi t/T)$,
 a three-dimensional (3D) plot of the energy spectrum for the two lowest-energy  Bloch bands is displayed in  Fig.~\ref{Fig1}(b), where $\varepsilon^{}_{0}$ is the energy modulation amplitude, $T$ is the cyclic  time of the pumping and $k$  is the wave vector in a unit of $1/(2L)$. It is seen in Fig.~\ref{Fig1}(b) that the energies of the two lowest-energy Bloch bands vary with time and are separated by a nonzero bandgap throughout the pumping.
Figure~\ref{Fig2}(a) shows the energy spectrum of a finite DQD chain  under the same modulation of  $V^{}_{0}(t)=\varepsilon^{}_{0}\sin(2\pi t/T )$ and $\delta V(t)=\varepsilon_{0}\cos(2\pi t/T)$. Here it is seen that two gapless bands are present in the Bloch band gap. Figure~\ref{Fig2}(b) shows the probability density distributions of two gapless band states at a selected energy. Here, it is seen that the two gapless band states are highly localized  at the ends of the DQD chain. Thus these gapless band states are of edge states as one commonly observes in a topological insulator.  Figure~\ref{Fig2}(c) shows the energy spectrum of the finite DQD chain under a trivial modulation of  $V^{}_{0}(t)=\varepsilon_{0}\sin(2\pi t/T)$, $\delta V(t)=\varepsilon_{\iota}+\varepsilon_{0}\cos(2\pi t/T)$ with the shifting energy $\varepsilon_{\iota}>\varepsilon_{0}$. It is evident that no gapless edge state is presented in this trivial modulation case. Here and hereafter, we assume that the periodic DQD chain is defined in an InAs nanowire  with    $m^{\ast}_{}=0.023m^{}_{0}$~\cite{Winkler2003} ($m^{}_{0}$ is the free electron mass), $2L=120$~nm, $2d=20$~nm, $\varepsilon^{}_{0}=6.0$~meV, and the sum of the two barrier potentials fixed as  $V^{}_{2}(t)+V^{}_{1}(t)=36$~meV, and the cyclic time $T$ is large enough to ensure the adiabaticity of pumping.

Based on the bulk-boundary correspondence~\cite{Hatsugai2016}, the presence of the gapless  edge  states is correlated to the nontrivial topology of the chain with periodic boundary conditions. By regarding the time as a virtual momentum axis orthogonal to the wire axis, the Berry curvature of a Bloch band in the  two-dimensional Brillouin zone can be introduced~\cite{Thouless1983,Wang2013},
\begin{align}
F(k,t)=i\left(\langle \partial_{k} u_{}|\partial_{t}u_{}\rangle-\langle \partial_{t} u_{}|\partial_{k}u_{}\rangle\right)\ ,
\end{align}
with   $|u(k,t)\rangle$ being the Bloch function below the energy gap, and   the nontrivial topology is quantified by a nonzero Chern number~\cite{Thouless1983,Wang2013},
\begin{align}
\mathcal{C}=\frac{1}{2\pi}\int^{T}_{0}\int^{\pi}_{-\pi} F(k,t)  dtdk\ .
\label{CN}
\end{align}
For the case of a nontrivial modulation as illustrated in the inset of Fig.~\ref{Fig2}(a),  the calculated  distribution  of the Berry curvature on the $k-t$ plane is displayed in Fig.~\ref{Fig2}(d). Numerical evaluation of Eq.~(\ref{CN}) performed as in Ref.~\onlinecite{Fukui2005} verifies the nonzero Chern number of $\mathcal{C}=1$, confirming that the DQD chain under the nontrivial modulation is in the topological nontrivial phase. In addition, based on the theory of topological polarization~\cite{Smith1993,Asboth2016},
$\mathcal{Q}(t)=\frac{1}{2\pi}\int^{t}_{0}\int^{\pi}_{-\pi}F(k,t)dtdk$ can be  interpreted
as the time-dependent charge polarization (in a unit of $-e$) and the topological charge pumping can be visualized by  the time-evolution of $\mathcal{Q}(t)$ in a cycle, just as  shown in the inset of Fig.~\ref{Fig2}(b), with $\mathcal{Q}(T)=1$ under this nontrivial modulation.

\section{Topological spin  pumping}~\label{SECIII}

\begin{figure}
\centering
\includegraphics[width=0.48\textwidth]{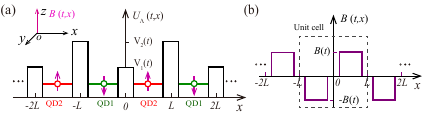}
\caption{(color online) (a) Schematic diagram of the DQD chain in the absence of QD well potential modulation but with the electron spin modulated by  a time-dependent staggered magnetic field $B(t,x)$ applied in the $z$ direction. (b) Schematic for the staggered external magnetic field  $B(t,x)$.     }
\label{Fig3}
\end{figure}

%with the arrows indicating the spin direction of each dot polarized by
%(a) The schematic diagram of the periodic double-quantum dot chain under a staggered magnetic field applied in the $z$ direction, without the plung. (b) The spatial distribution of the  time-dependent staggered magnetic field along the wire-axial direction.
In this section, the implementation of  topological  spin  pumping is proposed in the periodic DQD chain with  a time-dependent  staggered magnetic field $B(t,x)$ applied in a direction perpendicular to the wire axis, say the $z$ direction, as shown in Fig.~\ref{Fig3}(a). Here, we set the QD well potential unmodulated but keep modulation on the barrier potential difference  $\delta V(t)$. The Hamiltonian of the periodic DQD chain system  can now be written as
\begin{align}
H^{}_{\rm Z}=\frac{p^{2}}{2m^{\ast}_{}}+U^{}_{\rm \Lambda }(t,x)+\frac{g^{\ast}_{}\mu^{}_{\rm B}B(t,x)}{2}\sigma^{}_{z}\ ,
\label{HZ}
\end{align}
with $g^{\ast}_{}$ being the electron Land\'{e}  factor,  $\mu^{}_{\rm B }$ the Bohr magneton, and $\boldsymbol{\sigma}=(\sigma^{}_{x},\sigma^{}_{y},\sigma^{}_{z})$ representing the three Pauli matrices. Specifically, the time-dependent staggered magnetic field [see  Fig.~\ref{Fig3}(b)] within a unit cell of $-L\leq x<L$ takes the form of
\begin{align}
B(t,x)=B(t)\left[f^{+}_{1}(x)-f^{+}_{2}(x)\right] \ ,
\end{align}
with $B(t)$ denoting the  strength of the field and  $f^{+}_{1}(x)$ and $f^{+}_{2}(x)$ representing the same spatial functions as before.  It is important to note that the effect of the vector potential  within the Landau gauge, i.e, $\mathbf{A}=(-By,0,0)$, is ignored in Eq.~(\ref{HZ}), due to the one-dimensional nature of the system, i.e., an extremely strong confinement in the transverse  directions of the nanowire.

% based on the two lowest Zeeman-splitting energy levels on each dot, we can derive the discrete Hamiltonian  of the periodic  DQD chain under this circumstance.
Using the commutation relation $[H^{}_{\rm Z},\sigma^{}_{z}]=0$, the Hamiltonian  of  the periodic DQD chain in the second quantization form can be written as
\begin{align}
H^{}_{\rm Z,T}=H^{}_{\rm Z, \uparrow}+H^{}_{\rm Z,\downarrow}\ ,
\label{HZT}
\end{align}
with the two spin-polarized terms given by
\begin{align}
H^{}_{\rm Z, \chi=\uparrow/\downarrow}=&\sum^{}_{n} \left(t^{}_{{\rm in},0}a^{\dagger}_{n,\chi}b^{}_{n,\chi}+t^{}_{{\rm ex},0} a^{\dagger}_{n+1,\chi}b^{}_{n,\chi}+{\rm{h.c.}} \right) \nonumber\\
&~~~\mp \frac{\Delta^{}_{\rm z}}{2} \sum_{n}  \left(a^{\dagger}_{n,\chi}a^{}_{n,\chi}-b^{\dagger}_{n,\chi}b^{}_{n,\chi} \right).
\label{HST}
\end{align}
Here, the subscripts $\chi=\uparrow,\downarrow$ are spin indexes with the two spin states satisfying $\sigma^{}_{z}|\uparrow\rangle=|\uparrow\rangle$ and $\sigma^{}_{z}|\downarrow\rangle=-|\downarrow\rangle$,   $a^{\dagger}_{n,\chi}$ ($a^{ }_{n,\chi}$) and $b^{\dagger}_{n,\chi }$ ($b^{ }_{n,\chi }$) are  the  creation (annihilation) operators of the electron spin states   locating  in the two QDs in the $n$th cell,  and $\Delta^{}_{\rm z}(t)=g^{\ast}\mu^{}_{\rm B}B(t)$ is the time-dependent Zeeman splitting energy.

 \begin{figure}
\centering
\includegraphics[width=0.48\textwidth]{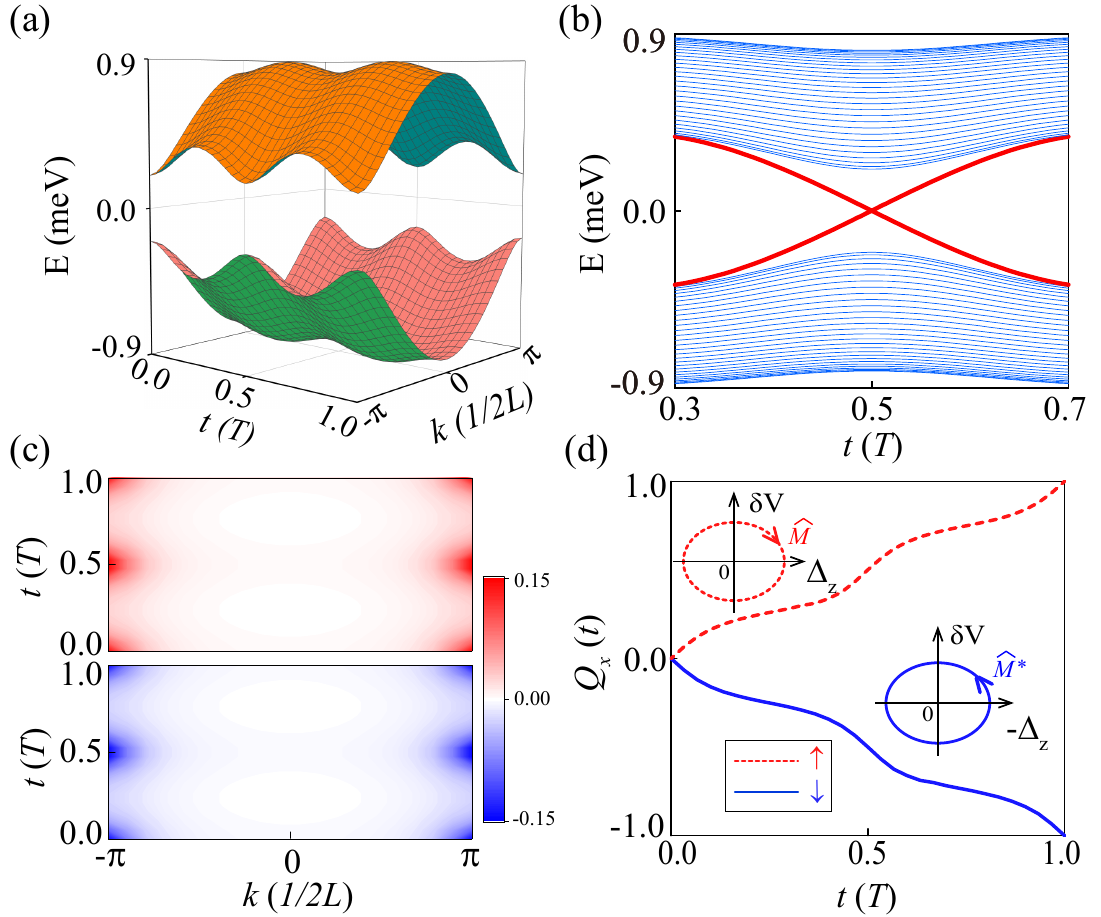}
\caption{(color online) (a) 3D plot of the energy spectrum of the Bloch bands for the periodic DQD chain  with the parameter vector $ \widehat{M}=(\Delta_{\rm z},\delta V)$ driven by a nontrivial modulation as shown in the inset of (d)  with   $\varepsilon^{}_{z}=0.9$~meV and $\varepsilon^{}_{0}=6.0$~meV.  (b) Energy spectrum of a correponding finite DQD chain of 30 unit cells versus $t$ under the nontrivial modulation. (c) Two spin-dependent Berry curvatures $F_{\chi=\uparrow,\downarrow}(k,t)$ calculated for the infinite DQD chain as a function of $k$ and $t$. The upper  and lower panels show the results for the  spin-up and spin-down Bloch bands.  (d) The time-evolutions  of the pumped spin polarizations $\mathcal{Q}_{\chi=\uparrow,\downarrow }(t)$ under this nontrivial modulation. The changing trajectories of the two parameter vectors  $ \widehat{M}=(\Delta_{\rm z},\delta V)$  and $ \widehat{M}^{\ast}=(-\Delta_{\rm z},\delta V)$  are shown in the inset with the arrows indicating the winding directions. The other parameters employed are the same as in Fig.~\ref{Fig1}.}
\label{Fig4}
\end{figure}

To achieve topological spin pumping in this system, we consider  a nontrivial modulation of  the parameter vector $\widehat{M}^{}_{ }=( \Delta^{}_{\rm z},\delta V)$ in a cycle. For the spin-up term, by mapping $ \Delta^{}_{\rm z}$ to $\Delta^{}_{0}$, the  Hamiltonian
 $H^{}_{\rm Z,\uparrow}$  in Eq.~(\ref{HST}) is equivalent  to the  Hamiltonian $H^{}_{0,T}$  in Eq.~(\ref{H0T}) and, therefore, the  spin-up   electron is  pumped  when the changing trajectory  of  $\widehat{M} $ has a nonzero winding number   in a cycle. The analysis can also be applied to the spin-down electron  with the parameter vector $\widehat{M}$ replaced  by $\widehat{M}^{\ast}=(-\Delta^{}_{\rm z},\delta V)$. Figure~\ref{Fig4}(a) shows a 3D plot of the  energy spectrum for the Bloch bands of the DQD chain under  a nontrivial modulation of $\Delta^{}_{\rm z}(t)=\varepsilon^{}_{z}\sin(2\pi t/T)$ and  $\delta V(t)=\varepsilon^{}_{0}\cos(2\pi t/T)$. Here, we see again that the energies of the two lowest-energy Bloch bands vary with time and are separated by a nonzero bandgap throughout the pumping. Note that, due to the spin degeneracy, each  energy band  shown in   Fig.~\ref{Fig4}(a) is twofold degenerate and comprised of two different spin states. Figure~\ref{Fig4}(b) shows the corresponding energy spectrum of a finite DQD chain under this nontrivial modulation. It is evident that two gapless edge-state Bloch bands  indicated by the  thick red curves are prresent in the bulk band gap.
Because the  modulation vectors $\widehat{M}^{}_{ }$ and $\widehat{M}^{\ast}_{ }$  have different winding directions in a cycle
[see the inset of Fig.~\ref{Fig4}(d)], the spin-up and spin-down electrons are therefore propagated in the  opposite directions and  results in the effective quantized  spin pumping.
%,  in which the gapless edge states is presented

To further illustrate this point, the Berry curvatures of the two  spin-polarized Bloch bands are introduced  $F^{}_{\chi =\uparrow,\downarrow}(k,t)=i\big(\langle \partial_{k} u_{\chi}|\partial_{t}u_{\chi}\rangle-\langle \partial_{t} u_{\chi}|\partial_{k}u_{\chi}\rangle\big)$, with $|u^{}_{\chi}(k,t)\rangle$  denoting the spin-dependent  Bloch functions below the energy gap.  Figure~\ref{Fig4}(c) shows  the  distributions of the two  Berry curvatures  on the $k-t$ plane under this nontrivial modulation. Figure~\ref{Fig4}(d) displays  the corresponding time-evolution  of the pumped spin polarization $\mathcal{Q}^{}_{\chi}(t)=\frac{1}{2\pi}\int^{t}_{0}\int^{\pi}_{-\pi}F^{}_{\chi}(k,t)dkdt$. The different signs in $\mathcal{Q}^{}_{\uparrow}(t)$ and $\mathcal{Q}^{}_{\downarrow}(t)$ implies the different propagating directions of the two  spin  polarizations~\cite{Schweizer2016}  and the topological spin pumping (in a unit of $-\hbar$) in a cycle becomes $ \frac{1}{2}\big[\mathcal{Q}^{}_{\uparrow}(T)-\mathcal{Q}^{}_{\downarrow}(T)\big]=1$. In addition, the equality $\mathcal{Q}^{}_{\uparrow}(T)+\mathcal{Q}^{}_{\downarrow}(T)=0$ demonstrates no net charge pumping in the whole process.

%when the parameter vector $\widehat{M}$ is subjected  ---

\section{Topological pumping in a finite DQD chain}~\label{SECIV}

\begin{figure}
\centering
\includegraphics[width=0.44\textwidth]{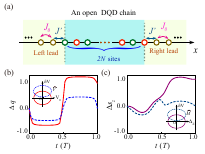}
\caption{(color online) (a) Schematic diagram of a finite DQD chain (comprising $2N$ sites) coupled to two external leads with $J^{}_{0}$ being the  inter-site coupling strength in the leads and  $J^{\prime}$ the strength of the tunneling coupling of the leads to the DQD chain.  (b) Transferred charge $\Delta q$, in a unit of $-e$, as a function of  $t$  with the parameter vector $\widehat{V}=(V_{0},\delta V)$ driven by a nontrivial and a trivial modulations as illustrated by the solid line and the dashed line in the inset of the panel.  (c) Transferred spin $\Delta s^{}_{z}$, in a unit  of $-\hbar$, versus  $t$ for the finite DQD chain with the parameter vector $\widehat{M}=(\Delta_{\rm z},\delta V)$ driven by a nontrivial and a trivial modulation as illustrated by the solid line and the dashed line in the inset of the panel.  Here, parameters $\varepsilon^{}_{z}=0.9$~meV,  $\varepsilon^{}_{0}=6.0$~meV, and $\varepsilon^{}_{\iota}=8.0$~meV are employed in the calculations and the other ones are the same as in Fig.~\ref{Fig1}.}
\label{Fig5}
\end{figure}

In this section, we consider  the case of a finite DQD chain coupled to two external leads, as illustrated in Fig.~\ref{Fig5}(a).
The electron energy spectrum of  the two semi-infinite leads in the momentum space can   be  derived as $E=2J^{}_{0}\cos(k)$, with $J^{}_{0}$ denoting   the  inter-site tunneling coupling strength in the leads and $J^{\prime}_{}$ being the strength of the tunneling coupling of the leads to the finite DQD chain (comprising $N$ cells or $2N$ sites). The eigenstates  of the leads are the degenerate   plane waves $\exp\left(\pm i kl\right)$~\cite{Aharony2002}, with the site index $l\leq 0$ for the left lead and $l\geq 2N+1$ for the right lead.
When the finite chain is subjected to an  incoming wave from the left side, the scattered solutions of the two leads takes  the form of~\cite{OEA2002a}
\begin{align}
|\psi^{}_{}\rangle=\begin{cases}
\exp(i k l )+\mathcal{R}(t)\exp(-i k l ) ~~~~~~l\leq 0\ , \\
\mathcal{T}(t)\exp(i k l )~~~~~~~~~~~~~~~~~~l\geq 2N+1\ ,
\end{cases}
\label{LFW}
\end{align}
with $\mathcal{T}(t)$ and $\mathcal{R}(t)$ representing  the time-dependent
transmission and reflection coefficients, respectively. Practically, the two time-dependent coefficients and  the transferred charge or spin  in a cycle can be  ascertained  by the second quantization Hamiltonian of the system based on the scattering matrix method~\cite{Brouwer1998,Zhou1999}.
%Based on the scattering matrix method,  the pumped charge or spin per cycle  actually depends  on the variation of the reflection coefficient  in a cycle

For a spinless chain, the second quantization Hamiltonian of the finite DQD chain  is given by Eq.~(\ref{H0T}) and the reflection coefficient can be  determined by the transfer matrix method (see Appendix). In terms of the reflection coefficient,  the transferred charge in a cycle can be expressed as
\begin{align}
\Delta q^{}_{}=\frac{e}{2\pi i} \int^{T}_{0} \frac{\mathcal{R}^{ }(t)}{dt}\mathcal{R}^{\ast}(t)dt \ .
\end{align}
For a spinful  DQD chain,  $ \boldsymbol{\mathcal{R}}(t)$ corresponds to a $2\times2$ matrix in spin space and  the transferred spin per cycle can be derived as~\cite{Meidan2010,Fu2006}
\begin{align}
\Delta \boldsymbol{\hat{s}}=\frac{\hbar}{2\pi i} \int^{T}_{0}{\mathrm Tr}\left[\frac{ \boldsymbol{\mathcal{R}}_{ }(t)}{dt}\boldsymbol{\hat{\sigma}} \boldsymbol{\mathcal{R}}^{\dagger}_{ }(t)\right] dt.
\end{align}
In the absence of spin-orbit interaction, $\boldsymbol{\mathcal{R}}(t)$ can be simplified to a diagonal form and the pumped spin is polarized by the external magnetic field along the $z$ direction.  Below,  we will demonstrate the  quantized charge and spin transports in a short DQD chain under the nontrivial modulation described above.

 Figure~\ref{Fig5}(b) shows the time evolutions of the transferred charge through a finite DQD chain  with $N=9$, $J^{}_{0}=0.4$~meV,  $J^{\prime}_{}=0.3$~meV,  $k\simeq\pm \pi/2$,  and  under the two  different modulations as illustrated in the inset of the figure.
Here, the effective charge pumping, i.e., $\Delta q=1$, is observed when the   parameter vector $\widehat{V}$ is subjected to a nontrivial modulation [see the red solid line]. The conclusion can also be applied to the spin pumping, but with the modulation vector $\widehat{V}$ replaced by $\widehat{M}=(\Delta_{\rm z},\delta V)$.  Figure~\ref{Fig5}(c) shows  the time evolutions of the transferred spin $\Delta s^{}_{z}$ under two different modulations. It is also evident that the transferred spin becomes quantized under a nontrivial modulation (see the purple solid line) and is zero under a trivial modulation  with  $ \Delta_{\rm z}(t)= \varepsilon_{z}\cos(2\pi t/T)$,
$\delta V(t)= \varepsilon^{}_{\iota} +\varepsilon_{0}\cos(2\pi t/T)$ and the shifting energy  $\varepsilon^{}_{\iota}>\varepsilon_{0}$ (see the blue dashed line).

% are solid and dashed lines represents the nontrivial modulation   and the dashed curve corresponds to a trivial modulation with $ \Delta_{\rm z}(t)= \varepsilon_{z}\cos(2\pi t/T)$, $\delta V(t)=\varepsilon^{}_{\iota} +\varepsilon_{0}\cos(2\pi t/T)$, $\varepsilon^{}_{z}=0.9$~meV, $\varepsilon^{}_{0}=6.0$~meV, and $\varepsilon^{}_{\iota}=8.0$~meV. ,  --- as illustrated by the two different changing trajectories of $\widehat{V}$ shown in the inset.

\section{Topological spin pumping in the presence of spin-orbit interaction}~\label{SECV}

It is well known that the spin-orbit interaction (SOI) plays an important role in the development of the topological insulator~\cite{Hasan2010,Qi2011,Kane2005,Bernevig2006}.
 In this section, we will show that in the presence of the Rashba SOI, the DQD chain can serve as a dynamic version of a topological insulator  and a spin pump if  the parameter vector $\widehat{M}=(\Delta^{}_{\rm z},\delta V)$ is driven by  a nontrivial modulation.

The Hamiltonian of the DQD chain in the presence of the staggered magnetic field and the Rashba SOI can be written as
\begin{align}
H^{}_{}=H^{}_{\rm Z}+\alpha p \sigma^{}_{y} \ ,
\label{OM-H}
\end{align}
where  $H^{}_{\rm Z}$ is to the Hamiltonian given in Eq.~(\ref{HZ}) and $\alpha$ represents the strength of the SOI, which is related to the spin-orbit length $x^{}_{\rm so}= \hbar/(m^{}_{\rm e}\alpha)$. For $x^{}_{\rm so}\gg L$,   the Hamiltonian of the DQD chain in the second quantization form can be derived as
 \begin{align}
 H^{}_{\rm T}=& \sum^{}_{n }\left(a^{\dagger}_{n } t^{}_{\rm in }b^{}_{n }+ a^{\dagger}_{n+1 }t^{}_{\rm ex }b^{}_{n } +\rm{h.c.}\right)\nonumber\\ &-\frac{\Delta_{\rm z}}{2} \sum^{}_{n}\left(a^{\dagger}_{n }\sigma^{}_{z} a^{ }_{n }-b^{\dagger}_{n }\sigma^{}_{z}b^{ }_{n }\right)\ ,
\label{DH-M}
 \end{align}
where $a^{}_{n}=\{a^{}_{n,\Uparrow},a^{}_{n,\Downarrow}\}$ and $b^{}_{n}=\{b^{}_{n,\Uparrow},b^{}_{n,\Downarrow}\}$ represent  the spinors constituted of   two spin-orbit states in the QDs and  the intra/inter-cell tunneling amplitudes can be  expressed as
\begin{align}
t^{}_{\rm in/ex}=t^{}_{\rm  in/ex,0}\exp[i\varphi^{}_{\rm in/ex}\sigma^{}_{x}]\ ,
\label{TM}
\end{align}
 with  $\varphi^{}_{\rm in/ex}$ being the functions of $L/x^{}_{\rm so}$. In fact, the phase factors appearing in the right side of Eq.~(\ref{TM}) arise purely from the interdot electron tunneling in the presence of the SOI~\cite{AC1984,Shahbazyan1994,Liu2021}.
\begin{figure}
\centering
\includegraphics[width=0.48\textwidth]{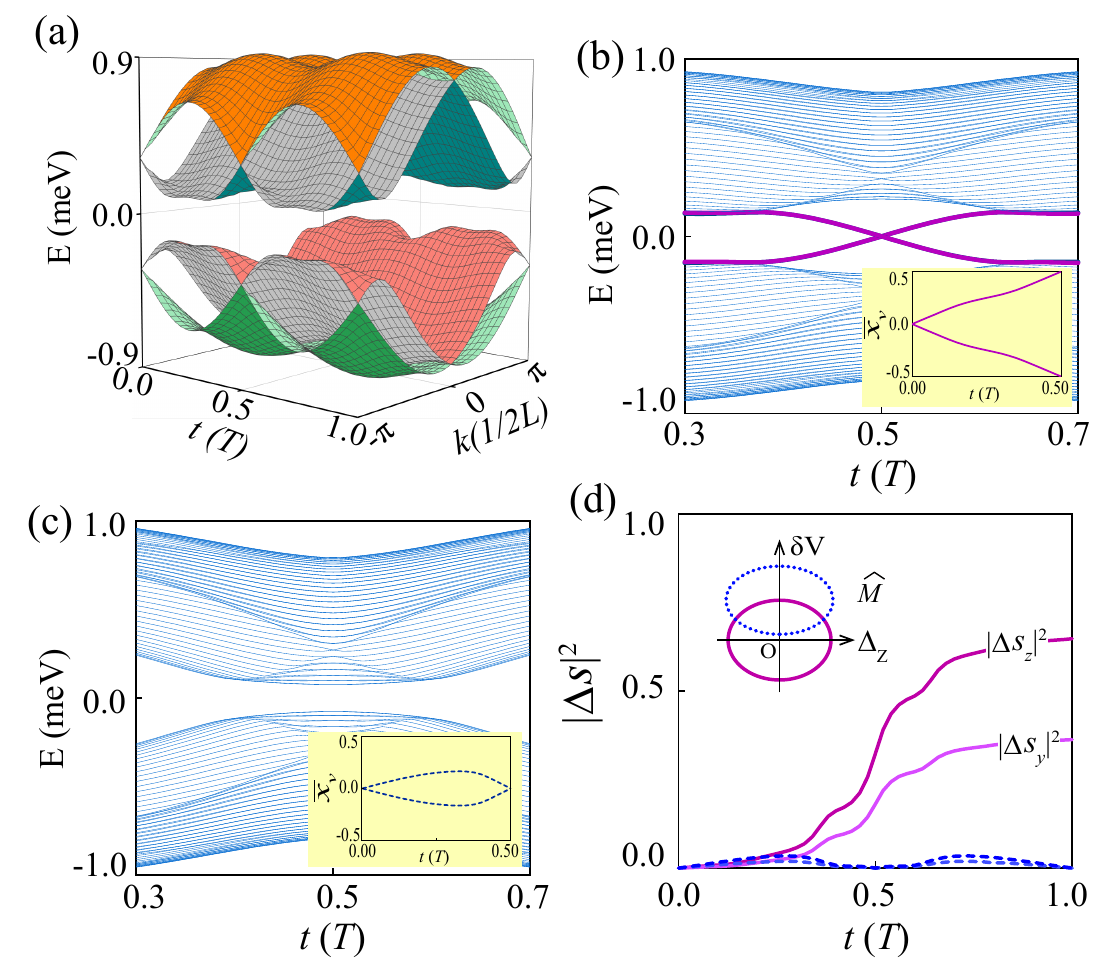}
\caption{(color online) (a) 3D plot of the energy spectrum for the four lowest-energy Bloch bands of the same DQD chain as in Figs.~\ref{Fig3} and \ref{Fig4}, but in the presence of the Rashba SOI with a spin-orbit length of $x^{}_{\rm so}=180$~nm, with the parameter vector $\widehat{M}=(\Delta_{z},\delta V)$ driven by a nontrivial modulation as shown by the solid line in the inset of (d) with $\varepsilon^{}_{  z} =0.9$~meV and $\varepsilon^{}_{ 0} =6.0$~meV. (b) Energy spectrum of a corresponding finite chain consisting of $30$ unit cells as a function of $t$ under the same nontrivial modulation.  Inset shows the time evolutions of the    charges on the two Wannier centers, $\bar{x}_{\nu=1,2}$, in a unit cell of the infinite DQD chain. (c) The same as (b), but for the finite DQD chain under a trivial modulation with $\varepsilon^{}_{  z} =0.9$~meV, $\varepsilon^{}_{ 0} =6.0$~meV and  $\varepsilon^{}_{ \iota}=8.0$~meV. (d) Transferred spin components $|\Delta s_{z}|^{2}$ (higher value line) and $|\Delta s_{y}|^{2}$ (lower value line) calculated for the finite DQD chain against $t$. The solid and dashed lines are the results of the calculations for the finite DQD chain with  the parameter vector $\widehat{M}$ driven by under the nontrivial (solid line) and trivial (dashed line) modulations as shown in the inset. All other undpecified parameters employed here are the same as in Fig.~\ref{Fig1}.}
\label{Fig6}
\end{figure}

Figure~\ref{Fig6}(a) shows a 3D plot of the energy spectrum for  the four lowest-energy Bloch bands of the DQD chain with $x^{}_{\rm so}=180$~nm~\cite{Scherubl2016} and with the parameter vector $\widehat{M}$ driven by the nontrivial modulation as illustrated by the solid line in the inset of Fig.~\ref{Fig6}(d). In contrast to the case without SOI, it indicates that the twofold degeneracy is generally lifted, except in the case of $t=0$, $T/2$, and $T$, for which the staggered magnetic field is zero and the system exhibits the Kramers' degeneracy. Figure~\ref{Fig6}(b) shows the energy spectrum of a corresponding finite chain. As expected, there exist two pairs of gapless edge-state Bloch bands crossing the band gap in the finite system under this nontrivial modulation. Figure~\ref{Fig6}(c) shows the energy spectrum of the finite DQD chain under a trivial modulation as illustrated by the blue dashed line in the inset of Fig.~\ref{Fig5}(d). Clearly, there is no  edge-state bands  in the band gap in this case.

% of the periodic chain
 In analogy with a topological insulator, the presence of the gapless  edge-state band in the band gap is  correlated with  the changing of time-reversal
polarization in a half cycle~\cite{Fu2006}.
To be more specific, this  kind of
  polarization is determined by the difference between   the  charges on the two Wannier centers in a unit cell constructed from the Bloch functions of the occupied bands~\cite{Soluyanov2011,Yu2011},
\begin{align}
\mathcal{P}(t)=\bar{x}^{}_{1}(t)-\bar{x}^{}_{2}(t)\ ,
\end{align}
where $\bar{x}^{}_{1}(t) $ and $\bar{x}^{}_{2}(t)$ represent the time-dependent charges on the Wannier centers in a unit cell with $x^{}_{\nu=1,2}(t) \in [-0.5,0.5]$.
The changing of time-reversal polarization in a half cycle can actually be identified as the (double-valued) $\mathbb{Z}_{2}$ invariant  and the gapless  edge  states only appear when the topological invariant is nonzero~\cite{Fu2006}.
The inset of Fig.~\ref{Fig6}(b) shows the time-evolutions of the  charges on the two Wannier centers in a unit cell of the infinite DQD chain under the nontrivial modulation. Indeed, it shows that the time-reversal polarization is increased by one in a half cycle, i.e.,  $\mathcal{P}(T/2)-\mathcal{P}(0)=1$, and  this is in contrast to a trivial modulation for which  $\mathcal{P}(T/2)-\mathcal{P}(0)=0$, see the inset of Fig.~\ref{Fig6}(c).
Therefore, the periodic DQD chain can  serve  as a  dynamical version of a topological insulator, with the topological property  determined   by   the changing trajectory of $\widehat{M}$.

 Similar to the topological spin pumping stated in  Sec.~\ref{SECIII}, there is no net charge pumped in a fully cycle and, because of the spin-orbit interaction,  the spin pumped per cycle is no longer with the quantization axis along the $z$ direction. However, the  results of the spin pumping  depends on whether the  changing  of the parameter vector $\widehat{M}$ is in a trivial or nontrivial trajectory in  a cycle. Figure~\ref{Fig6}(d) shows the time evolutions of the transferred spin in a finite DQD chain with the spin-orbit length $x_{\rm so}=180$~nm and  the vector  $\widehat{M}$ driven by the two different modulations as illustrated in the inset of the figure.  It is evident that the spin transferred per cycle is nonzero under the nontrivial modulation (see the solid lines). Here, we would like to note that for this particular case the pumped spin has both $z$ and $y$ components. It is also clearly seen that the spin transferred in a cycle is zero  under a trivial modulation (see the  dashed curves).

\section{Conclusion}~\label{SECVI}

 In this paper, we propose a scheme to  implement   adiabatic topological pumping in a  semiconductor nanowire  DQD chain. The topological property of the pumping is related to the changing trajectory of the modulation parameters. We show that the topological charge pumping can be achieved  by periodically modulating the QD well and   barrier  potentials, simultaneously, and the quantized charge  transfer can only be realized if the corresponding changing contour is characterized with a nonzero winding number in  a cycle.  When the QD well potential is replaced by  a time-dependent staggered magnetic field, the  topological  spin pumping  can be achieved by  a nontrivial modulation of the barrier potentials and magnetic field. We, in addition, demonstrate that in the presence of  the Rashba SOI, the periodic DQD chain can  serve as a dynamic version of a topological insulator and as  a spin pump under a nontrivial modulation of the  barrier potentials and magnetic field.
Even though  the topological adiabatic pumping  studied in this paper  is based on  an InAs nanowire, it can also be extended to other 1D nanostructures, such as  carbon nanotubes and  Ge/Si heterostructure nanowires~\cite{Biercuk2005,Hu2007},  with long  spin-relaxation times.
 Our theoretical study presented in this work should boost the exploration of adiabatic topological pumping and  higher-dimensional topological phases of matter in one-dimensional semiconductor nanostructures.

\begin{center}
\textbf{ACKNOWLEDGMENTS}
\end{center}

This work is supported by the Ministry of Science and Technology of China through the National Key Research and Development Program of China (Grant Nos. 2017YFA0303304 and 2016YFA0300601), the National Natural Science Foundation of China (Grant Nos. 91221202, 91421303, and 11874071), the Beijing Academy of Quantum Information Sciences (No. Y18G22), and the Key-Area Research and Development Program of Guangdong Province (Grant No. 2020B0303060001).\\

\begin{center}
\textbf{CONFLICT OF INTEREST}
\end{center}

The authors have no conflicts to disclose.

\begin{center}
\textbf{DATA AVAILABILITY}
\end{center}

The data that support the findings of this study are available
from the corresponding author upon reasonable request.

\begin{appendix}

\setcounter{equation}{0}

\renewcommand\theequation{A\arabic{equation}}

\section*{Appendix: DERIVATION OF THE REFLECTION COEFFICIENT}

In this appendix, the detailed derivation for the reflection coefficient $\mathcal{R}(t)$ of a finite DQD chain coupled to two external leads is given.

Based on the discrete model shown  in Fig.~\ref{Fig5}(a), the   Hamiltonian of the total system can be written in the second quantization form as
\begin{align}
H^{}_{\rm tot}=H^{}_{\rm DQD}+H^{}_{\rm L}+H^{}_{\rm R}+H^{}_{\rm C} \ ,
\label{THG}
\end{align}
where $H^{}_{\rm DQD}$ is the  Hamiltonian of the finite DQD chain, $H^{}_{\rm L/R}$ the  Hamiltonian of the lead locating in the  left/right side of the chain,  and  $H^{}_{\rm C}$    describes  the coupling between the leads and the DQD chain. For a finite DQD chain comprised by $N$ DQD cells,  the Hamiltonian can be explicitly written as
   \begin{align}
H^{}_{\rm DQD}=&\sum^{N}_{n=1 } \left(t^{}_{\rm in,0}a^{\dagger}_{n}b^{}_{n}+t^{}_{\rm ex,0} \xi_{n} a^{\dagger}_{n+1}b^{}_{n}+\rm{h.c.} \right) -\frac{\Delta^{}_{0}}{2} \nonumber\\
 &\times\sum^{N}_{n=1} \left(a^{\dagger}_{n}a^{}_{n}-b^{\dagger}_{n}b^{}_{n} \right) \ ,
  \end{align}
  with  $ \xi_{n}$ depending on the cell index $n$, which equals to 1 for $1\leq n<N$ and $0$ for $n=N$.
The Hamiltonian of  the two external leads are given by
\begin{align}
H^{}_{\rm L }=&J_{0} \sum^{ }_{l \leq -1}\left(c^{\dagger}_{l}c^{}_{l+1}+{\rm h.c.}\right)\ ,  \nonumber\\
H^{}_{\rm R}=&J_{0} \sum^{}_{l\geq 2N+1}\left(c^{\dagger}_{l}c^{}_{l+1}+{\rm h.c.}\right)\ ,
\label{LHS}
\end{align}
with $c^{\dagger}_{l}$ ($c^{ }_{l}$)  representing the  electron  creation (annihilation) operator  on the $l$-th site of the  leads  and  $J^{}_{0}$ denoting  the  inter-site tunneling strength. The coupling of the two leads to the DQD chain can be described as
\begin{align}
H^{}_{\rm C}=J^{\prime}_{}\left(a^{\dagger}_{1}c_{0}+b^{\dagger}_{N}c_{2N+1}+\rm{h.c.}\right)\ ,
  \end{align}
  with  $J^{\prime}$ denoting the strength of the tunneling couplings at the two interfaces.

On the basis of Eqs.~(\ref{LHS}), the electron energy spectrum of  the two semi-infinite leads in the momentum space  can be  derived as $E=2J^{}_{0}\cos(k)$ and the corresponding eigenfunctions are given in the forms of the plane waves $\exp\left(\pm i kl\right)$~\cite{Aharony2002}.
 When the DQD chain is subjected to a right-moving input wave from the left lead, the scattering solution   can be written as Eq.~(\ref{LFW}). By exploiting the scattering matrix formalism, the input wave is connected to the reflected wave  by the transfer equation~\cite{Meidan2010}
\begin{align}
\mathcal{T}(t)
\begin{pmatrix}
e^{ik(2N+2)}\\
e^{ik(2N+1)}
\end{pmatrix}={\bf\mathcal{V}}\begin{pmatrix}
1+\mathcal{R}(t)\\
e^{-ik}+\mathcal{R}(t)e^{ik }
\end{pmatrix}\ ,
\label{TRSE}
\end{align}
with $\mathcal{T}(t)$ and $\mathcal{R}(t)$ representing the  transmission and reflection coefficients of the DQD chain, respectively, and the transfer matrix $\mathcal{V}$ can be obtained from
\begin{align}
\mathcal{V}=\mathcal{O}\mathcal{P}_{}^{N}\mathcal{I}\ .
\label{TSM}
\end{align}
Here, $\mathcal{I}$ represents the matrix corresponding to the input transition,
$\begin{pmatrix}
1+\mathcal{R}(t)\\
e^{-ik  }+\mathcal{R}(t)e^{ik  }
\end{pmatrix}
\longrightarrow
\begin{pmatrix}
B_{1}\\
A_{1}
\end{pmatrix}
$, with $A_{n}$ and $B_{n}$ ($n=1,2,...,N$) denoting the wave function amplitudes on the QD1 and QD2 of the $n$-th unit cell, respectively,
and is given by \begin{align}
\mathcal{I}=\begin{pmatrix}
\frac{2E+\Delta^{}_{0}}{2t^{}_{\rm in,0 }}&-\frac{J^{\prime}}{t^{}_{\rm in,0 }}\\
1&0
\end{pmatrix}\begin{pmatrix}
\frac{E}{J^{\prime}}&-\frac{J_{0}}{J^{\prime}}\\
1&0
\end{pmatrix}.
\label{IMAT}
\end{align}
$\mathcal{P}$ represents the transfer matrix within the DQD chain $
\begin{pmatrix}
B_{n}\\
A_{n}
\end{pmatrix}\longrightarrow\begin{pmatrix}
B_{n+1}\\
A_{n+1}
\end{pmatrix}
$, and has the form of
\begin{align}
\mathcal{P}=\begin{pmatrix}
\frac{2E-\Delta_{0} }{2t^{}_{\rm  ex,0}}&-\frac{t^{}_{\rm in,0 }}{t^{}_{\rm  ex,0}}\\
1&0
\end{pmatrix}\begin{pmatrix}
\frac{2E+\Delta_{0} }{2t^{}_{\rm  in,0}}&-\frac{t^{}_{\rm ex,0 }}{t^{}_{\rm in,0}}\\
1&0
\end{pmatrix}.\label{PMAT}
 \end{align}
 $\mathcal{Q}$ is the output transition matrix,
\begin{align}
\mathcal{O}=
\begin{pmatrix}
\frac{E }{J_{0}}&-\frac{J^{\prime}}{J^{}_{0}}\\
1&0
\end{pmatrix}\begin{pmatrix}
\frac{2E-\Delta^{}_{0}}{2J^{\prime}_{ }}&-\frac{t^{}_{\rm in,0 }}{J^{\prime}}\\
1&0
\end{pmatrix}.\label{OMAT}
\end{align}
By substituting Eqs.~(\ref{IMAT})-(\ref{OMAT}) into Eq.~(\ref{TSM}),  we can derive the explicit form of the transfer matrix $\mathbf{\mathcal{V}}$ and find the  reflection coefficient by exploiting   Eq.~(\ref{TRSE}),
\begin{align}
\mathcal{R}(t)=\frac{[\mathcal{V}]_{2,2}e^{2ik}+[\mathcal{V}]_{2,1}e^{ik}-[\mathcal{V}]_{1,2}e^{ik}-[\mathcal{V}]_{1,1}}{[\mathcal{V}]_{1,1}-[\mathcal{V}]_{2,2}+[\mathcal{V}]_{1,2}e^{-ik}-[\mathcal{V}]_{2,1}e^{ik}}\ ,
\end{align}
with  $[\mathcal{V}]^{}_{n,m}$ ($n,m=1,2$) representing the matrix elements of the transfer matrix.

\end{appendix}

\end{document}